# SOUTH POL: Revealing the Polarized Southern Sky

A. M. Magalhães[a], C. M. de Oliveira[a], A. Carciofi[a], R. Costa[a], E. M. G. Dal Pino[a], M. Diaz[a], T. Ferrari[a], C. Fernandez[a], A. L. Gomes[a], L. Marrara[a], A. Pereyra[c], N. L. Ribeiro[a], C. V. Rodrigues[b], M. S. Rubinho[a], D. B. Seriacopi[a], and K. Taylor[a]

[a]*Instituto de Astronomia, Geofísica e Ciências Atmosféricas, Universidade de São Paulo, Rua do Matão 1226, São Paulo, SP 05508-090, Brazil*
[b]*Instituto Nacional de Pesquisas Espaciais/MCT, Av. dos Astronautas 1758, São José dos Campos, SP 12227-010, Brazil*
[c]*Observatório Nacional, Rua Gal. José Cristino 77, Rio de Janeiro, RJ 20921-400, Brazil*

**Abstract.** SOUTH POL will be a survey of the Southern sky in optical polarized light. It will use a newly designed polarimetric module at an 80cm Robotic Telescope. Telescope and polarimeter will be installed at CTIO, Chile, in late 2012. The initial goal is to cover the sky south of declination -15° in two years of observing time, aiming at a polarimetric accuracy $\lesssim$ 0.1% down to V=15, with a camera covering a field of about 2.0 square degrees. SOUTH POL will impact areas such as Cosmology, Extragalactic Astronomy, Interstellar Medium of the Galaxy and Magellanic Clouds, Star Formation, Stellar Envelopes, Stellar explosions and Solar System, among others.

## MOTIVATION

Several astrophysical processes polarize (linearly and/or circularly) the incoming light beam, either at the source or along the way to the observer. Therefore we stand to gain a lot of information, encoded in the polarization state of the light, by measuring this polarization. Examples of such processes in UV/optical/near-infrared (NIR) domain (and places they might appear) are:

Dust scattering: Interstellar Medium (ISM); Young Stellar Objects; AGB stars

Thompson scattering: Cosmic Microwave Background (CMB); hot star envelopes

Synchrotron radiation: Active Galactic Nuclei (AGN) and AGN hot spots; Gamma-ray Bursts (GRBs)

Cyclotron radiation: Magnetic cataclysmic binaries (polars).

Despite the increase in the use of optical/NIR imaging polarimetry and spectropolarimetry in the past few decades [1, 2], and in contrast with the radio/sub-mm domain, **there has been no single, broad band, all-sky polarimetric survey in the optical/NIR yet, either ground- or space-based**.

And yet, there is an ever increasing need for a polarimetric survey in the optical/NIR domain. The current scientific race to detect B-mode polarization in the CMB radiation [3] is a case in point. That detection and its interpretation depend crucially on the subtraction of the Galactic foreground emission. O*ptical* dust polarization provides the best mask for the foreground position angle [4]. Unfortunately, as noted by [3] and [4], the optical Galactic polarization is not known well enough, especially away from the Galactic Plane, precisely where the foreground correction is needed the most.

At the same time, knowledge of the optical polarization of celestial sources is of paramount importance in several key astrophysical areas. We outline some of these areas in the next few sections. They provide the motivation for **SOUTH POL,** a survey of the polarized sky at optical wavelengths with an 80cm Robotic Telescope (RT), the latter to be installed at CTIO in late 2012.

## INSTRUMENTATION

An imaging polarimeter is being designed that will use a rotating half-wave plate, followed by either a calcite Savart plate or polarizer and a broad band filter [5]. The detector will be a 9K x 9K, 92 mm wide square EEV CCD, which will provide a 2.0 $deg^2$ field of view at the f/4.5 80cm RT. Due to the relatively large physical size of the optics required, the waveplate and prism will each consist of four individual pieces mounted on a mosaic. The polarimeter will be built as a module that can be easily swapped with the imaging camera planned for the RT. We aim at covering the sky south of declination -15°.in about 2 yrs of observing time.

## SOUTH POL STRATEGY

One of the main motivators for SOUTH POL is observing the interstellar polarization essentially all over the southern sky (Fig. 1), especially away from the Galactic Plane. The needed accuracy has to be also such that the data in general may attain the widest impact possible in other areas, such as stellar envelopes and AGN, while also being achievable in a reasonable amount of observing time.

High Galactic Latitude clouds (HGLC) will be among the ISM targets presenting the biggest challenge. Our experience with the HGLC observed for the Southern Interstellar Polarization Survey [6] shows that one needs to get down to V~15-16 to map a typical cloud. Here, $A_V \sim 0.3$, typically. This will in turn originate a typical polarization $P_V \lesssim 3A_V \sim 9E_{B-V} \sim 0.1\%-1.0\%$. Hence, an accuracy of $P/\sigma_P = 5$, ensuring a positive detection, implies $\sigma_P \sim 0.1\%$ for $P \sim 0.5\%$. For V=15, this will require 5 min per waveplate position, at eight (8) positions. Including overheads, each such field will require about 50 min for completion. Accuracies for a range of V magnitudes are given in **Table 1**. They were estimated assuming a 22 mag/arcsec$^2$ sky, air mass 1, readout noise of 10 $e^-$ and 8 waveplate positions.

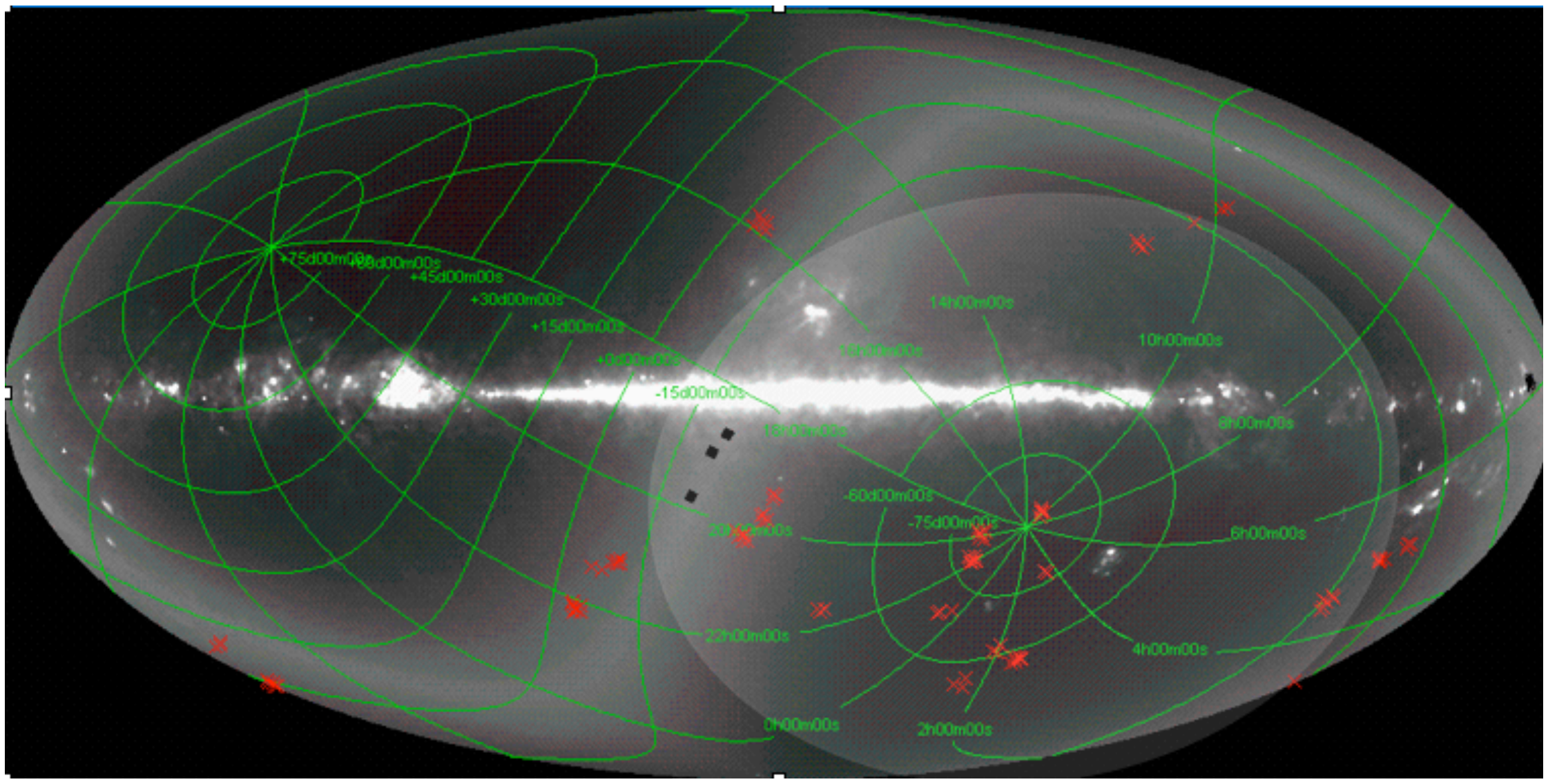

**FIGURE 1.** Region of the sky (lighter area) to be initially observed by SOUTH POL, i.e., Dec $\leq$ -15°. It includes the Magellanic Clouds, most of the Southern Milky Way and the Galactic Center. Crosses indicate High Galactic Latitude Clouds observed in our current survey (Magalhães et al. 2005).

**TABLE 1.** SOUTH POL accuracies (%)

| V (mag) | 8 × 60 sec | 8 × 300 sec |
|---|---|---|
| 13 | 0.088 | 0.039 |
| 14 | 0.140 | 0.062 |
| 15 | 0.223 | 0.100 |
| 16 | 0.361 | 0.160 |
| 17 | 0.600 | 0.263 |
| 18 | 1.051 | 0.449 |
| 19 | 2.011 | 0.83 |

Using the stellar population synthesis model of the Galaxy provided by the Besançon group [7], we find that V=15 mag stars (of any spectral type) observed towards Galactic latitude -90° will span about 3.5kpc, easily encompassing (and hence sampling) the Galactic dust layer. A total of about 90 stars in the bin 15<V<16 are expected per degree (or 180 stars per CCD field) at that latitude.

## SOUTH POL'S LEGACY

Expected contributions from the Survey include:

**Cosmology**: Good sampling of the interstellar polarization will allow better modeling of the Galactic foreground polarization for CMB studies by past, current and future sub-mm telescope and missions.

**Extragalactic Astrophysics**: An accuracy ≤ 1.0% at V~19 (Table 1) will allow many blazars to be detected. Identification of CGRO/EGRET and FERMI sources will be feasible. The magnetic field structure of the Magellanic Clouds and other nearby systems will be amenable to study.

**Galaxy, ISM and Star Formation**: Testing of proposed grain alignment mechanisms, such as radiative torque [8] will be possible. SOUTH POL data, in conjunction with stellar parallaxes from GAIA or LSST, will provide an unprecedented knowledge of the 3-D magnetic field structure of the Galaxy. Also, the magnetic field topoly across molecular clouds from low $A_V$ (optical) through high $A_V$ (sub-mm) will be known.

**Stellar Astrophysics**: Follow-up of GRBs and SNe will provide statistics and time evolution of explosive phenomena. Also, the orientation of stellar envelopes with regards to the ambient magnetic field [9] may be studied in more detail.

In summary, SOUTH POL will be an unprecedented undertaking in Astronomy. The first planned epoch (~2 yr of observing) of the survey will provide polarization of stellar or point sources (i.e., sources not larger than 14 arcsec) south of declination -15°. The survey will progress to more northernly declinations, as more time becomes available with the RT.

Its legacy will have an immediate impact in several areas, as exemplified above. Subsequent epochs of SOUTH POL will uncover the polarimetric variability of both galactic & extragalactic sources, as well as study the polarization of extended sources. Future and more detailed, pointed, deeper programs with the RT and the imaging polarimeter will additionally contribute to tackle more specific problems.

## ACKNOWLEDGMENTS

SOUTH POL has been funded by the São Paulo funding agency, FAPESP, through grant no. 2010/19694-4 (PI: AMM). AMM is partly funded by CNPq. The Robotic Telescope has been funded by FAPESP (grant no. 2009/54202-8, PI: CMO).